\documentclass[12pt]{article}
\usepackage{amsmath,amssymb,amsthm,mathrsfs}
\usepackage{array}
\usepackage{graphicx,psfrag,epsf}
\usepackage{multirow}
\usepackage{multicol}
\usepackage{amssymb}
\usepackage{mathtools}
\usepackage{booktabs}
\usepackage{dsfont}
\usepackage{appendix}
\usepackage{float}
\usepackage[colorlinks,citecolor=blue,linkcolor=blue,urlcolor=blue]{hyperref}
\usepackage{algorithm,algpseudocode}
\usepackage[bottom=0.5in,top=1.5in]{geometry}

\newcommand{\blind}{1}

\newcommand{\bigo}{\mathcal{O}}
\newtheorem{theorem}{Theorem}
\newtheorem{lemma}{Lemma}

\newtheorem{remark}{Remark}

\DeclareMathOperator*{\argmax}{\arg\!\max}

\begin{document}

\def\spacingset#1{\renewcommand{\baselinestretch}%
{#1}\small\normalsize} \spacingset{0.95}


\if1\blind
{
  \title{\bf Data-guided Treatment Recommendation with Feature Scores}
  \author{Zhongyuan Chen, Ziyi Wang, Qifan Song, and Jun Xie \\ \\
    {\it Department of Statistics, Purdue University} \\
     {\it 150 N. University Street, West Lafayette, IN 47907} \\
   junxie@purdue.edu 
    }
  \date{\today}
  \maketitle
} \fi

\if0\blind
{
  \bigskip
  \bigskip
  \bigskip
  \begin{center}
    {\LARGE\bf Data-guided Treatment Recommendation with Feature Scores}
\end{center}
  \medskip
} \fi

\spacingset{1.5} 

\begin{abstract}
Despite the availability of large amounts of genomics data, medical
treatment recommendations have not successfully used them.
In this paper, we consider the utility of high dimensional 
genomic-clinical data and nonparametric methods for making cancer treatment
recommendations. This builds upon the framework of the individualized  
treatment rule \cite{Qian11} but we aim to overcome their method's
limitations, specifically in the instances when the method encounters
a large number of covariates  
and an issue of model misspecification. We tackle this problem using a
dimension reduction method, namely Sliced Inverse Regression (SIR, \cite{Li91}),
with a rich class of models for the treatment response. Notably,
SIR defines a feature space for high-dimensional data,
offering an advantage similar to those found in the popular 
neural network models. With the features obtained from SIR, a simple
visualization is used to compare 
different treatment options and present the recommended
treatment. Additionally, we derive the  
consistency and the convergence rate of the
proposed recommendation approach through a value function. 
The effectiveness of the proposed approach is demonstrated through
simulation studies and the promising results from a real-data example
of the treatment of multiple myeloma.
\end{abstract}

\vfill


\section{Introduction}
Conventionally, the approach to recommending disease treatments has been
through expert-driven 
guidance, based on knowledge built over decades. With the
availability of large 
amounts of data, there is a growing interest in using 
data to help choose different treatment options. For instance, extensive 
amounts of genomics data have 
been generated in cancer research, e.g., genetic mutations, mRNA
expression, along with 
clinical data including treatment options and clinical outcomes. These
data add 
valuable information to support and complement expert knowledge for
cancer treatments. In this
paper, we aim to develop a data-guided 
tool with simple 
visualizations that will help doctors and patients evaluate
different treatment options and make treatment recommendations. 

As a case study, we examine a data set of gene expressions and
treatment responses of multi-center clinical trials of 
bortezomib in treatment of multiple myeloma
\cite{Mulligan07}. Multiple myeloma is a malignant bone marrow
cancer. This disease is highly heterogeneous, meaning that different 
patients with diverse genomic information show different clinical
outcomes \cite{Mitra17}. However, the current treatment
strategy is limited to the experience of physicians and experts,
mainly using
patient's clinical information such as age and cancer stage. With the specific  
genomic-clinical data set, we aim to
make a treatment recommendation 
between two therapeutic choices, a traditional chemotherapy named
dexamethasone and a targeted drug bortezomib.

Our goal is related to research on
precision medicine, which has attracted a considerable amount of
interests.  A recent study   
on precision oncology for acute myeloid leukemia \cite{Gerstung17}
analyzed genomic-clinical data to support clinical
decision-making. \cite{ZhuXie15} used a nonparametric method to
identify patient subpopulations that would experience stronger treatment
effects than the rest of the patient population. These studies,
however,  were exploratory with
no formal framework to define an optimal treatment rule. A valuable
formulation has been contributed by \cite{Qian11}.
Formally, we consider a list of random variables  
$(\mathbf X, A, Y)$ from a genomic-clincial dataset, where $Y$ denotes
a treatment response variable (the larger value the better), $\mathbf X \in
{\cal X} \subset \mathbb{R}^p$ 
denotes a set of clinical covariates plus
genetic variables, e.g., gene expressions, and
$A \in \cal{A}$ denotes the treatment 
index taking values in a finite discrete space of treatment options,
e.g., ${\cal A}=\{-1,1\}$ corresponding to treatment and control,
or ${\cal A}=\{1,...,M\}$ corresponding to $M$ treatment
options. A treatment
recommendation rule is a function $d(\mathbf X): {\cal X} \to {\cal
  A}$. It is called an individualized treatment rule in
\cite{Qian11}. An individualized
treatment rule that gives the highest mean response is the optimal
one that we hope to find.

There are two types of approaches to construct the optimal
treatment rule. One is refered to as direct methods
and the other, indirect methods. The direct methods include Outcome
Weighted Learning \cite{Zhao12}, Residual Weighted Learning
\cite{Zhou17}, and other variational forms \cite{Dasgupta2020}. The
basic idea is to directly optimize a criterion, called the Value
function, through the support vector machine (SVM) 
machine learning method. These 
approaches, however, are often confined by the limitation of the SVM
procedure, e.g., the difficulty with a small separation margin, choices of
kernels, etc. The indirect methods consist of two steps. The first step
is to estimate
a conditional mean of treatment response given 
clinical variables and the treatment index, $E(Y|\mathbf X,
A)$. The optimal treatment 
rule is then defined as the one that maximizes the estimated conditional mean 
\cite{Rosenwald02, van08, Qian11,
  Cui17, Hager18, Bai17, Zhao19}. There are also methods focusing on
dynamic treatment regimes that consider treatment recommendations
at multiple times as well as Bayesian approaches of dynamic treatment regimes
\cite{Schulte14, Zhang13, Daniel19, Liu18, Yang18,
Laber17, Xu16, Murray18}. The indirect methods rely heavily on the
correct model specification for the conditional mean $E(Y|\mathbf X,
A)$ \cite{Qian11}, which is often challenging to achieve. In
addition, none of the existing methods 
are good at handling high-dimensional data.

We focus on the indirect method and improve it by developing an
approach that contains a class of rich
conditional mean models. More
specifically, we apply Sliced Inverse
Regression, or SIR \cite{Li91}, to predict the treatment 
response. SIR is designed to retrieve
interesting features of high-dimensional data by low-dimensional
projections. The method is able 
to model the relationship between a treatment response and a set of
genomic and clinical variables through an arbitrary unknown
function. There is no linear model
assumption about the conditional mean of the treatment
response. Instead, the model space of the SIR method is often bigger
than other indirect methods. We also present the theory to show that the richer
model space of SIR leads to improved treatment recommendation. 

An important strength of the SIR procedure is that it directly
estimates the low-dimensional projection space and represents the
high-dimensional data by a few features. This resembles the
feature definition component of the neural network models that are popularly used
nowadays. We name the SIR projected data Feature Scores. Specifically,
SIR will work well in the instance when there is no strong effect from an 
individual clinical or genetic variable but the treatment response may
depend on
an unknown feature, which is a commonly occuring situation in cancer treatment. A simple
scatter plot of the treatment 
response versus Feature Score allows users to visualize and compare
different treatment options. Thus, our method offers a tool for doctors and
even patients to assess and confirm the available treatment
plans. Moreover, we prove that
the SIR procedure consistently estimates the optimal treatment rule
under moderate assumptions.

In summary, the biggest contribution of this article is to define a
small feature space in the framework of individualized treatment
rules. The major advantages of the proposed method include 1) dimension
reduction with feature detection, 2) rich conditional mean models 
for consistent estimation of the optimal treatment, 3)
visualization of the optimal treatment
recommendation, 4) theoretical guarantee with a convergence rate. 

The remainder of the article is organized as follows. In Section 2, we
introduce the value function, define the Feature Score, and show a
visualization of the treatment recommendation. In Section 3, we prove the
consistency and derive the convergence rate of the proposed
recommendation approach. In Section 4, we show simulations and compare
our proposed method with other methods. Section 5
demonstrates the results of applying the proposed method to the case study
of treatments for multiple myeloma. Some discussions are given
in Section 6. The Supplementary Materials include the information of
data and code and technical proofs of the lemma and thereom.

\section{Treatment recommendation through Feature Scores}
Formally, we have a set of random variables  
$(\mathbf X, A, Y)$ in the data set, where $\mathbf X \in {\cal X}
\subset \mathbb{R}^p$ 
denotes clinical covariates plus a big set of genetic variables, 
$A \in {\cal A}$ is the treatment 
index taking values in a finite discrete space $\cal{A}$ of treatment options,
$Y$ is the treatment response variable with larger values
indicating better treatment response. A treatment
recommendation rule is a function $d(\mathbf X)$ with
values in the space of $\cal{A}$. Denote the
distribution of $(\mathbf X, A, Y)$ by $P$, and the conditional mean
$E(Y|\mathbf X, A)$. Following the framework of
individualized treatment rules \cite{Qian11}, we will first show an
optimal treatment recommendation rule must maximize $E(Y|\mathbf X,
A=a)$ over $a \in \cal{A}$. This result justifies the indirect
methods, which focus on the estimation of $E(Y|\mathbf X, A=a)$. Next,
we will apply SIR \cite{Li91} to estimate $E(Y|\mathbf X, A=a)$ and
then obtain the optimal recommendation rule. The model space for
estimation of $E(Y|\mathbf X, A=a)$ in the SIR method is very large,
which is the biggest advantage of our proposed method.

\subsection{Value function and optimal recommendation}
By convention, we use upper case letters for random variables and
lower case letters for values of the random variables. 
The likelihood of $(\mathbf X, A,
Y)$ under $P$ is $f_0(\mathbf x)p(a|\mathbf x)f_1(y|\mathbf x,a)$, where  $f_0$ is
the unknown density of $\mathbf X$, $p(\cdot|\mathbf x)$ is the randomization probability
of $A$ given $\mathbf X=\mathbf x$, and $f_1$ is the unknown distribution of $Y$
conditional on $(\mathbf X, A)$. Let $P^d$ denote the distribution of $(\mathbf X, A,
Y)$ when a treatment recommendation rule $d(\mathbf X)$ is used to assign
treatments, then the 
likelihood becomes $f_0(\mathbf x)\mathds{1}(d(\mathbf x)=a)f_1(y|\mathbf x,a)$. Define the
Value of $d$ as $V(d) \triangleq E^{d}(Y)$.
Assume $p(a|\mathbf x) > 0$ for any $a\in \mathcal{A}$ and $\mathbf x\in
\mathcal{X}$. The Value of any treatment rule $d$ can be
expressed as   
\[
    V(d) = \int Y dP^d = \int Y \frac{dP^d}{dP} dP =  \int Y \frac{\mathds{1}_{d(\mathbf X)=A}}{p(A|\mathbf X)}dP = E\left[Y \frac{\mathds{1}_{d(\mathbf X)=A}}{p(A|\mathbf X)}\right]
\]
An optimal treatment recommendation rule, denoted as $d_0$, is a rule
that has the maximum Value 
over all possible treatment recommendation rules, 
\[
    d_0 \in \argmax_{d} V(d). 
\]
Moreover, denote $ Q_0(\mathbf X,A)\triangleq E(Y|\mathbf X, A) $. We also have
\[
    V(d) = E\left[\frac{\mathds{1}(d(\mathbf X)=A)}{p(A|\mathbf X)} E[Y|\mathbf X,A]\right] = E\left[ \sum\limits_{a\in \mathcal{A}}\mathds{1}_{d(\mathbf X)=a}Q_0(\mathbf X,a)\right] = E\left[Q_0(\mathbf X,  d(\mathbf X))\right].
\]
Note the Value for the optimal treatment rule $V(d_0) = E[Q_0(\mathbf X,
  d_0(\mathbf X))] \leq E[\max\limits_{a\in\mathcal{A}} Q_0(\mathbf X,
  a)]$. Meanwhile by the definition of $d_0$, $V(d_0)
\geq V(d)|_{d(\mathbf X) \in \argmax_{a\in\mathcal{A}} Q_0(\mathbf X, a)} =
E[\max\limits_{a\in \mathcal{A} }Q_0(\mathbf X, a)] $. Thus, the
optimal treatment rule satisfies $d_0(\mathbf X) \in
\argmax\limits_{a\in\mathcal{A}}Q_0(\mathbf X,a)$. Our goal is to estimate
$d_0$, which will be achieved by first estimating the conditional mean
$ Q_0(\mathbf X,A)$.

More specifically,  the estimated treatment
recommendation rule is defined as
\begin{equation}\label{rule_est}
  d(\mathbf X) \in
  \argmax\limits_{a \in \mathcal{A}}Q(\mathbf X, a),
\end{equation}
where $Q(\mathbf X, A)$ is an estimator of the 
true conditional mean $Q_0(\mathbf X,A)$. The following result,
modified from \cite{Qian11}, shows that the difference between the
largest Value $V(d_0)$ and $V(d)$ is controlled by the mean squared
error of the estimator $Q(\mathbf X,A)$.

We require an assumption similar to the margin condition in
classification. Assume both the true conditional mean $Q_0(\mathbf X,A)$ and
its estimator $Q(\mathbf X, A)$ are square integrable.
Define $T(\mathbf X,A) = Q(\mathbf X,A)-E[Q(\mathbf X,A)|\mathbf X]$ and $T_0(\mathbf X, A) =
Q_0(\mathbf X,A)-E[Q_0(\mathbf X,A)|\mathbf X]$. They are referred to as the treatment effect
terms in \cite{Qian11}. The following assumption is about the margin
of $T_0$, i.e., the difference in mean responses between the optimal
treatment and the suboptimal treatment.
\begin{enumerate}
\item[(A.1)] There exist some constants $C > 0$ and $\alpha > 0$ such
  that
  \[\textbf{P}\left(\max\limits_{a\in\mathcal{A}}T_0(\mathbf X,a) -
    \max\limits_{a\in\mathcal{A}\setminus \argmax_a T_0(\mathbf X,a)} 
    T_0(\mathbf X,a)\leq \epsilon\right) \leq C\epsilon^{\alpha}
\]
for  any $\epsilon > 0$.
\end{enumerate}

\begin{lemma}
 Suppose $p(a|\mathbf x) \geq S^{-1}$ for a positive constant $S$ for all
 $(\mathbf x,a)$ pairs and assume (A.1). For any treatment rule $d:
 \mathcal{X} \mapsto  \mathcal{A} $ and square integrable function $Q:
 \mathcal{X} \times \mathcal{A} \mapsto \mathbb{R}$ such that
 $d(\mathbf X) \in \argmax_{a\in \mathcal{A}} Q(\mathbf X, a)$, we have  
 \[
     V(d_0) - V(d) \leq C^{'}\left[E(Q(\mathbf X,A)-Q_0(\mathbf X,A))^2\right]^{(1+\alpha)/(2+\alpha)}
 \]
 where $C^{'} = (2^{2+3\alpha}S^{1+\alpha}C)^{1/(2+\alpha)}$. 
 \label{thm1}
\end{lemma}
The proof is in the Supplementary Materials.

\subsection{A rich conditional mean model}
Lemma 1 justifies the use of the indirect methods. When we have
a consistent 
estimator of $Q_0(\mathbf X, A)=E(Y|\mathbf X, A)$, that is, an
estimator $Q(\mathbf X, A)$ converges to $Q_0(\mathbf X, A)$, Lemma 1
shows the Value of the estimated treatment
recommendation rule, i.e., $V(d)$ of $d(\mathbf X) \in
\argmax\limits_{a \in \mathcal{A}}Q(\mathbf X, a)$, will also converge to the
optimal value $V(d_0)$. However, this will not happen if the
conditional mean is modeled incorrectly. 
In fact, if the approximation space used in estimating $Q_0$
does not contain the truth, then the estimated treatment recommendation
rule will not be consistent. \cite{Qian11}
pointed out this challenge but did not present methods to address it.
We attempt to offer a solution via Sliced Inverse Regression (SIR)
\cite{Li91}. SIR  
is a novel method for reducing the dimension of $\mathbf X$ without going
through any model-fitting 
process in the first place. It is developed under a very general model,
$Y=g(\beta_1\mathbf X, \beta_2\mathbf X,...,\beta_k\mathbf X, \epsilon)$,
where $\beta$'s are unknown row vectors, $k$ is a small number,
$\epsilon$ is the error term  
independent of $\mathbf X$, and $g$ is an arbitrary unknown
function. Applying SIR, we make a very general assumption:
\begin{enumerate}
\item[(A.2)] For each treatment group $a \in
\cal{A}$, the conditional mean response depends on a low-dimensional
  projection of $\mathbf X$. That is, $E[Y|\mathbf X, A=a]=E[Y|\beta_{a,1}\mathbf X,
  \beta_{a,2}\mathbf X,...,\beta_{a,k}\mathbf X, A=a]$, with $k$ as a
  small number, e.g., $k=1$ or 2. 
\end{enumerate}
In other words, given 
treatment $a \in \cal{A}$, the conditional mean response is assumed
$E(Y|\mathbf X, A=a)=\eta_a(\beta_{a,1}\mathbf X,
\beta_{a,2}\mathbf X,...,\beta_{a,k}\mathbf X)$, where $\eta_a$ is an
arbitrary and unknown function. 
The projection of a set of predictors
$\mathbf X$ onto the $k$ dimensional subspace, $(\beta_{a,1}\mathbf X,
\beta_{a,2}\mathbf X,...,\beta_{a,k}\mathbf X)$,
captures all we need to know about $Y$ for the given treatment
$A=a$. The projection space and the arbitrary function $\eta_a$ are
allowed to be different for different treatment groups $a \in
\cal{A}$. This assumption 
offers a rich class of models for the conditional mean $Q_0(\mathbf X,
A)$. Specifically, if we denote $\cal{Q}$ as the approximation space
for $Q_0$, then $\cal{Q}$ contains the linear model, the commonly used
generalized linear models, and many more, because the link function
$\eta_a$ can take any functional form. This offers a much richer class
of individualized treatment rules than the other existing methods.

The number $k$
is supposed to be very small, e.g., 1 or 2, and $\beta_{a,1}\mathbf X$, or
$(\beta_{a,1}\mathbf X, \beta_{a,2}\mathbf X)$, provides
summary information
of a patient for prediction of the treatment response. We name
$\beta_{a,1}\mathbf X$, or $(\beta_{a,1}\mathbf X, \beta_{a,2}\mathbf X)$
if $k=2$, Feature Score. The use of the Feature Score has the
advantage of representing the cancer
treatment situation where there 
would be no strong effect from an individual genetic variable
but the treatment response would depend on  
unknown features. \cite{Li91} provided a direct etimator of
$\beta$'s through the SIR procedure. For
each treatment group $A=a$, suppose we have patient
samples of the treatment response and the covariate vector $\{(y_i,
\mathbf x_i)\}$. We apply SIR and obtain the first projection
direction $\hat{\beta}_1$. The Feature Score is denoted as
$u_i=\hat{\beta}_1 \mathbf x_i$, which can be interpreted as a summary 
feature of a patient and is supposed to capture the majority data
information for the prediction of the treatment response $Y$. More
interestingly, this feature definition is analogous to that of the
neural network model, and SIR is able to directly estimate the features
without knowing the link function $\eta_a$.

\subsection{Simple visualization with Feature Score}
Suppose ${\cal A}=\{1,...,M\}$, so there are $M$ different treatment
groups in a given data set. We conduct SIR for each treatment
group and obtain the first projection direction $\hat{\beta}_{a,1}$,
$a=1,...,M$. We can project
all patients onto a one-dimensional space (line) and calculate
their Feature Scores  
$u_i=\hat{\beta}_{a,1} \mathbf x_i$, where the Feature Scores will be
different for different treatment groups. We draw a simple scatter plot of
$y_i$ versus $u_i$ for each treatment group,  
$a=1,...,M$. Even though the Feature Scores of different treatment
groups $u_i=\hat{\beta}_{a,1} \mathbf x_i$ are not comparable to each
other, we can still compare the treatment response via the vertical
axis, which has the same scale over different scatter plots (see
Figure \ref{fig:obs_scatter}). These plots provide 
visualization of the treatment options, i.e., $a=1,...,M$, where
larger vertical values indicate better treatment response.

We also obtain a nonparametric fitting of the function,
$\hat{g}_a(u)$, for example, by local constant estimates, or
LOESS (locally weighted smoothing) \cite{Cleveland88}, for each of
the treatment groups $a=1,...,M$. These nonparametric estimates
provide the predicted treatment responses for each treatment option.
Given a new patient with data vector $\mathbf x$, we first
calculate its Feature Score, $u_{a}=\hat{\beta}_{a,1}\mathbf x$, then decide
its treatment option to be the one
maximizing the predicted treatment responses. More specifically, we
will recommend a treatment choice as
\begin{equation}\label{plan}
\argmax_{a=1,...,M} \tilde{g}_{a}(\hat{\beta}_{a,1}\mathbf x),
\end{equation}
where $\hat{\beta}_{a,1}$ is from the SIR procedure and $\tilde{g}_{a}(\cdot)$
is the nonparametric function estimate based on the patient samples
$\{(y_i, u_i)\}$ with $u_i=\hat{\beta}_{a,1} \mathbf x_i$.  
In general, we can have the subspace dimension $k>1$ and SIR may
project data of different treatment groups onto different
subspaces. Nevertheless, we will obtain a nonparametric estimate of
the functional relationship, $\tilde{g}_{a}(\hat{\beta}_{a,1}\mathbf x,
..., \hat{\beta}_{a,k}\mathbf x)$. The treatment recommendation will be
similarly defined as (\ref{plan}).

\begin{figure}[ptbh]
\centering
\includegraphics[width=1.0\linewidth]{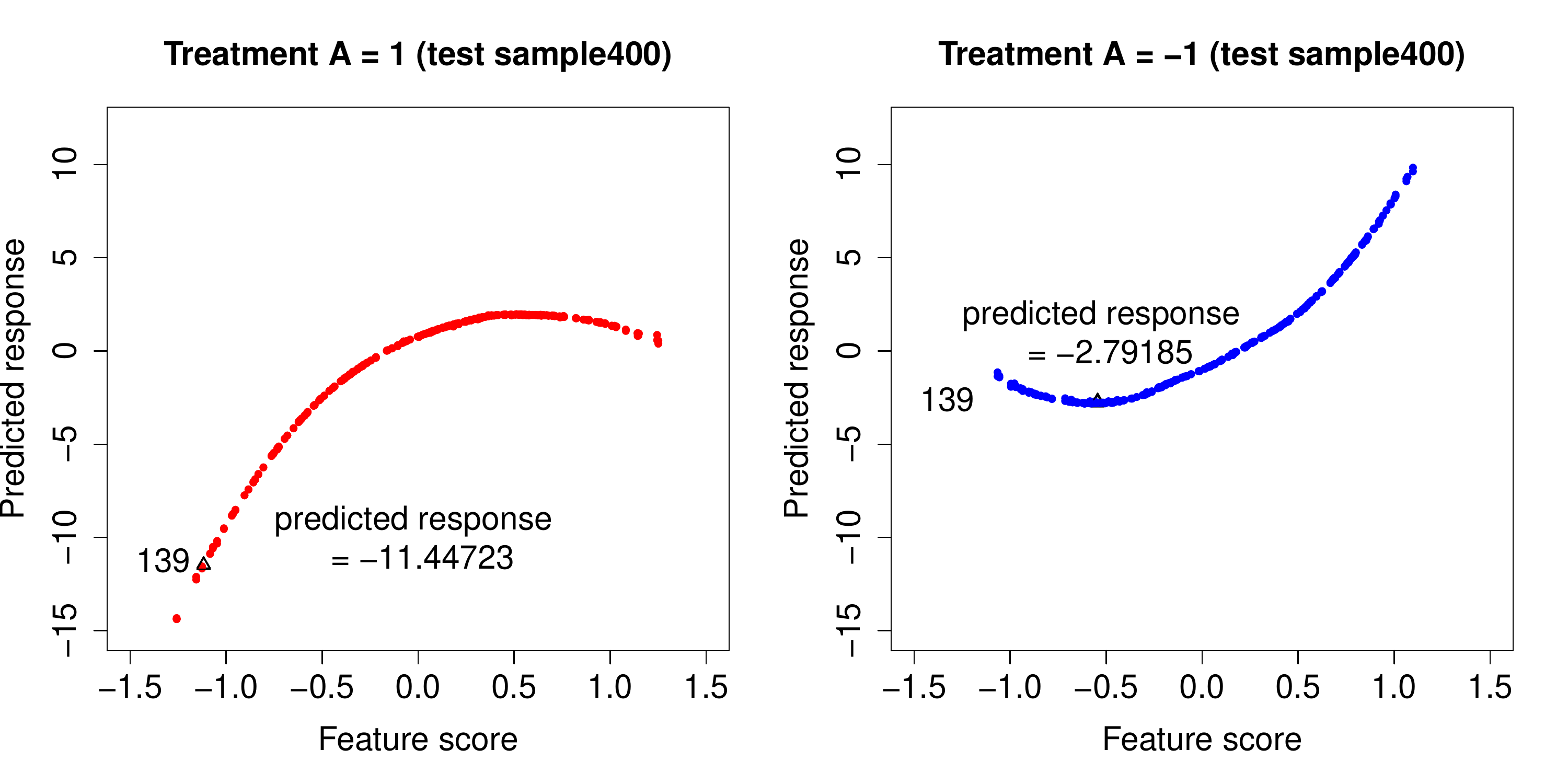}
\caption{\small{Predicted response value versus Feature
  Score $u_{a}=\hat{\beta}_{a}\mathbf{x}$ under each treatment (Scenario 3). A specific data point, ID 139, is marked for visualization of the optimal treatment.}}\label{fig:simplot}
\end{figure}

The visualization through the scatter plot of $y_i$ versus Feature Score
$u_i$ is a very useful tool. For a patient with Feature
Score $u_a=\hat{\beta}_{a,1}\mathbf x$, we can locate it on the horizontal
axis (or the projected space when the Feature Score is more than one
dimensional) and then look at treatment response values  
based on the vertical 
axis in the scatter plots, as shown in Figure \ref{fig:simplot}. We can also compare the predicted
treatment responses between our proposed treatment plan and the plan 
based on current expert guidelines, according to the vertical axis. This
will show what improvements in 
treatment response may
be achieved from the proposed treatment recommendation.

\subsection{Data preprocessing and the algorithm}
Before implementing the SIR procedure, we should go through a few
steps of data preprocessing. The first step is to confirm that a given
genomics data set contains significant information for the prediction
of treatment response. We evaluate the overall dataset information through a
global hypothesis testing method, the Cauchy combination test 
developed by \cite{Liu19}. The p-value from the Cauchy combination
test serves as evidence to support data-guided treatment
recommendations. If the Cauchy combination test gives a large p-value,
we should not consider the genomic data in forecasting a patient's
prognosis and for recommending treatments.

The second step of data preprocessing is to conduct initial variable
selection before implementing SIR when we analyze a large number of
genomic variables. SIR is a dimension reduction method 
involving principle component analysis (PCA). In general, some initial
reduction in dimensionality is desirable before applying any PCA-type
methods \cite{Johnstone09}. Although more recent developments of sparse 
SIR \cite{Lin21, Lin19} may be directly applied, we instead consider
here two variable 
selection methods and incorporate selection into the SIR process. 
One method is to select variables with the smallest 
p-values from a simple regression of $Y$ over
$X_j$ and $A$, $j=1,...,p$, at a false discovery rate (FDR) cutoff, e.g.,
5\%. Another is to  
screen for important variables from nonparametric local regression of  $Y$ over
$X_j$ and $A$ using LOESS, with the smallest 5\%
residual errors. A user can choose to
use either variable selection method before implementing SIR.

To determine the number of Feature Scores $k$, which is the dimensions
for reduction in the proposed SIR model, we can use the $\chi^2$ test
suggested by 
\cite{Li91}. On the other hand, as SIR is a PCA-type method, it is 
common practice to consider one or two Feature Scores, i.e., one or
two principal components, for visualization. The specific algorithm of
our treatment recommendation is provided in the table below. 

\begin{algorithm}
[!ptbh]\caption{\it Treatment recommendation procedure}\label{alg:treat}

\begin{algorithmic}[1]\Procedure{$s=$ \sf
    TreatRcmd}{$Y,\mathbf{X}, A, \mathbf{x}_{new}$} 
\begin{description}
\item \textit{Input}: A training data set with observed $(\mathbf{X},A,Y)$, where $Y$ is the treatment response, $\mathbf{X}=
  (X_1,\ldots,X_p)$ is the set of genomic variables and clinical
  covariates, and $A$ is the treatment index; A new observation
  with vector value $\mathbf{x}_{new}$ for treatment recommendation.

\item \textit{Output}: Scatter plots of $Y$ versus Feature Scores;
  The predicted response under each treatment option for $\mathbf{x}_{new}$
   and the optimal treatment option.
\end{description} 

\textbf{\!\!$\triangleright$ Overall information summary}
\State Calculate p-value from the Cauchy combination test.
\State Alert if the overall p-value is large. Continue only if the p-value is small.

\textbf{\!\!$\triangleright$ Subset selection (Optional)}
\State Select a subset of $X_j$'s for the following SIR procedure,
using either linear regression or nonparametric local regression of $Y$ over 
$X_j$ and $A$. A default
cutoff is the false discovery rate 5\%, or using LOESS with the 
 smallest 5\% residual errors. 

\textbf{\!\!$\triangleright$ Dimension reduction (SIR)}
\State For each treatment group $A=a$, conduct SIR to obtain the
low-dimensional projection directions 
$\hat{\beta}_a$.
\State Make scatter plots of $Y$ versus Feature Score
$u_a=\hat{\beta}_a\mathbf{X}$ for each treatment group.

\textbf{\!\!$\triangleright$ Prediction}
\State For the new data point $\mathbf{x}_{new}$, calculate its Feature
Scores $u_a=\hat{\beta}_a\mathbf{x}_{new}$ under each
treatment option $A=a$ and predict the response
under the corresponding treatment.
\State Obtain the optimal treatment recommendation that gives the
largest predicted response.

\EndProcedure
\end{algorithmic}
\end{algorithm}

\section{Consistency and convergence rate}
Our treatment recommendation rule is $d(\mathbf X) \in \argmax\limits_{a \in
  \mathcal{A}} Q(\mathbf X, a)$, where $Q(\mathbf X, A)$ is an estimator of $Q_0(\mathbf X,
A)=E(Y|\mathbf X, A)$ and is obtained by SIR and the nonparametric procedure
LOESS. Recall the Value function defined in Section 2.1. The following theorem
shows we can have $V(d)$ converging
to the optimal Value $V(d_0)$ with a certain rate. Besides the margin
condition (A.1), we require
additional assumptions from SIR \cite{Li91} and for the nonparametric LOESS
estimator. We first rewrite the SIR assumption (A.2) by denoting the
treatment index as $i 
\in {\cal A} =\{1,\dots, M\}$ and the projection directions $\beta$'s
as $\mathbf
B_i \in \mathbb{R}^{k \times p},  k < p$. 

\begin{enumerate}
\item[(A.2)]\label{assumdim} There exist some
full-rank matrices $\mathbf B_i \in \mathbb{R}^{k \times p},  k < p$,
such that $E[Y|\mathbf X, A=i] = E[Y|\mathbf B_i \mathbf X,
A=i]=\eta_{i}(\mathbf B_i \mathbf X)$, where $\eta_i(\cdot)$'s are
$\rho$-Lipschitz continuous and have continuous second derivatives.  
Furthermore, for any row vector $\mathbf \xi \in \mathbb R^{p}$, $E\left[ \xi \mathbf
  X| \mathbf{B}_i \mathbf X\right]$ is a linear function of
$\mathbf{B}_i \mathbf X$. Besides, the
dimension of the central inverse curve $E\left[\mathbf X| y,
  A=i\right]$ equals to the dimension of the space spanned by the
columns of $\mathbf B_i$, $col(\mathbf B_i)$, and the variance
$v_i(u) = Var[Y | \mathbf{B}_i \mathbf X = u, A=i]$
is a continuous function. 

\item[(A.3)]\label{kernelassum} Denote the kernel function of LOESS by
  $K_H(u) = 
  |H|^{-1/2}K(H^{-1/2}u)$, where $u \in \mathbb R^{k}$ and the bandwidth matrix
  $H \in \mathbb R^{k \times k}$. Assume the kernel function $K(\cdot)$ is
  $\rho$-Lipschitz, compactly supported, and satisfies $\int
  uu^{\top} K(u)du = \mu_2(K)\mathbf I$, where $\mathbf I$ is the
  identity matrix and $\mu_2(K)$ is a constant depending on
  $K$. Moreover, all odd-order moments of $K$ equal to 
  zero, that is, $\int u_1^{l_1}\cdots u_d^{l_d}K(u)du = 0$ for all
  non-negative $l_1\cdots l_d$ when their sum is odd. Additionally,
  the bandwidth matrix $H$ is symmetric and positive definite with each
  entry, as well as $n^{-1}|H|$, tending to 0 as $n \to \infty$, and
  the ratio of the largest and the smallest eigenvalue of $H$ is uniformly
  bounded for all $n$. 
  
\item[(A.4)]\label{dfassump} For all $i \in {\cal A}$, let
  $f_i(\cdot)$ be the conditional density function of
  $\mathbf{B}_i\mathbf{X}$ given $A=i$. Assume that $f_i(\cdot)$ is
  uniformly bounded away from 0 and has a continuous gradient
  function $D_{f_i}(\cdot)$.
\item[(A.5)] Denote
  $n_i=|{\{j: A_j=i\}}|$ as the number of observations in the
  treatment group $A=i$. Assume $\min_{i\in\mathcal A}P(A=i)>c$ for some positive
  constant $c$ and the support set of $\mathbf X$ is bounded. 
\end{enumerate}

As represented in (\ref{plan}) in Section 2.3, we write the treatment
recommendation rule as 
$d(\mathbf x) \in 
\argmax\limits_{i \in {\cal A}} Q(\mathbf x, i)$, where $Q(\mathbf x,i) =
\tilde g_{i}(\widehat{\mathbf B_i}\mathbf x)$ with $\widehat{\mathbf
  B_{i}}$ as the estimated projection directions from SIR and
$\tilde{g}_{i}(\cdot)$ the LOESS function from the 
  training data 
  $\{\widehat{\mathbf B_{i}}\mathbf  x_j, y_j\}_{\{j: A_j=i\}}$.

\begin{theorem}
Assume (A.1)-(A.5). The difference
between the optimal Value, $V(d_0)$, and $V(d)$ of our treatment
recommendation rule converges to $0$ in probability as $n \to \infty$:
\begin{equation}
\begin{aligned}
V(d_0)-V(d) &\leq \left(|H|^{-1} \|H^{-1/2}\|_F^2 \bigo_p(\frac{1}{n})+\bigo_p\left(\frac{|H|^{-1/2}}{n}+ \|H\|_1^2\right)\right)^{\frac{1+\alpha}{2+\alpha}},
\end{aligned}
\label{them2eq}
\end{equation}
where  $\|H\|_1$ denotes the maximum column absolute sum 
  and $\|\cdot\|_F^2$ denotes the Frobenius norm.
When the bandwidth matrix $H=diag\{h,\cdots, h\}$ with $h =
n^{-\frac{1}{k+3}}$, the upper bound on the right hand side becomes
$\bigo_p(n^{-\frac{2(1+\alpha)}{(k+3)(2+\alpha)} })$. 
\label{thm2}
\end{theorem}

The proof is in the Supplementary Materials.


\begin{remark}
 Theorem \ref{thm2} is obtained by combining the estimation errors of the SIR
 procedure and the LOESS nonparametric regression. The second error
 term in (\ref{them2eq}) is the intrinsic estimation error of the
 LOESS regression and the 
 first error term is the additional estimation error induced by the
 uncertainty of the SIR procedure.  
  
\end{remark}

\begin{remark}
The conditional mean model assumption and the linearity condition in
(A.2) are from SIR. 
Note that $\mathbf{B}_i$ and $\eta_i(\cdot)$ are not identifiable
(e.g., one can always multiply $\mathbf{B}_i$ to any $k \times k$
full-rank matrix), but the space spanned by the columns of
$\mathbf{B}_i$ is unique.
\end{remark}

\begin{remark}
The smoothness assumption of $\eta_i$ and $v_i$,
the requirements on the kernel choice in
(A.3), along with (A.4),
ensure the consistency of the nonparametric estimation for each mean
regression function $\eta_i$ via the local linear regression
approach. The bandwith matrix $H$ usually takes simple form as
$diag\{h, \cdots, h\}$, where $h > 0$. Given this simplification
the last statement in assumption (A.3) is automatically satisfied. 
\end{remark}

\begin{remark}
The compactness assumption on the support set of $\mathbf X$ in (A.5)
greatly facilitates our theoretical analysis, for example, it
trivially ensures that $\|D_f(\cdot)\|$ is bounded. This assumption is
reasonable for most medical treatment applications, since the patient
measurements, such as gene expression levels, are usually bounded or
standardized. We conjecture that our theoretical results will still
hold for unbounded $\mathbf X$ such as the Gaussian design, while the
rigorous convergence analysis for such cases is left for future
studies.   Assumption (A.5) also ensures that $n_i\asymp n$ in
probability.
\end{remark}

\begin{remark}
For the simplicity of representation, our theorem only
considers the fixed $p$ and $k$ situation. 
If $p$ and $k$ increase with respect to $n$, then the corresponding
convergence rates can be rigorously studied by utilizing the high
dimensional algorithm and theory of SIR developed by, e.g.,
\cite{Zhu2006} and \cite{Lin18}, \cite{Lin21}, \cite{Lin19}. In
Section 3 in the Supplmentary Materials, we present a convergence result under
$p\rightarrow\infty$ and $p/n\rightarrow 0$. 
\end{remark}

\section{Simulation studies}
To assess the proposed method, we perform extensive simulations. We
compare our method with several existing approaches, including Outcome
Weighted Learning (OWL) \cite{Zhao12}, Residual Weighted Learning (RWL)
\cite{Zhou17}, and a linear regression method with ordinary least
squares estimation of the
conditional mean of the treatment response, denoted as OLS. 

We generate $p$ covariates $X_1,...,X_p$ from uniform
$[-1,1]$, where a small and a large covariate set are considered with
$p=8$ or 100. We consider two treatment options ${\cal A}=\{1, -1\}$
of a randomized controlled study. The response $Y$ follows a normal
distribution with mean $\mu(\mathbf{x})+t_0(\mathbf{x})a$ and standard deviation 1, where
$\mu(\mathbf{x})$ represents the effect of the covariates
$\mathbf{x}=(x_1,x_2,\ldots,x_p)$ 
and $t_0(\mathbf{x}) a$
represents the treatment effect, which may depend on $\mathbf{x}$. We simulate two sample sizes $n=100$ and
$n=400$, with half of the samples in the treatment group and the other
half in the control group. The
terms $\mu(\mathbf{x})$ and $t_0(\mathbf{x})a$ are chosen from the following four
scenarios:
\begin{enumerate}
\item $\mu(\mathbf{x}) = 2+4x_1+4x_2+4x_3$, when $a=1$;\\
  $\mu(\mathbf{x}) = (2+4x_1+4x_2+4x_3)^2$, when $a=-1$; \\
  $t_0 = 0$.

\item $\mu(\mathbf{x}) = 2+2x_1+2x_2+4x_3+4x_4$; $t_0(\mathbf{x}) = 1.3(x_2-2x_1^2+0.3)$.

\item $\mu(\mathbf{x})=\frac{10x_1}{0.5+(x_2+1.5)^2}$; $t_0 (\mathbf{x})= 1.3(x_2-2x_1^2+0.3)$.

\item $\mu(\mathbf{x}) = \frac{10x_1}{0.5+(x_2+1.5)^2}$; $t_0 (\mathbf{x})= 3.8(0.8-x_1^2-x_2^2)$.
\end{enumerate}
Scenario 1 is modified from a simulation model of OWL
\cite{Zhao12}. We define the mean function as a linear  
function for  $a = 1$ and its quadratic function
for  $a=-1$. Scenario 2 is similar to the second scenario
in RWL \cite{Zhou17}. Scenario 3 and 4 have 
nonlinear functions with $\mu(\mathbf{x})$ 
modified from a simulation model of SIR \cite{Li91}. 

We first conduct the Cauchy combination test
on the entire set of 
covariates $X_1,...,X_{p}$. The Cauchy combination test gives 
siginificant results for all simulations. For data sets with a small number
of covariates, i.e., $p=8$, we directly implement the SIR method. On
the other hand, for data sets where the dimension is comparable to the
sample size, i.e., $p=100$, we conduct initial
variable selection before implementing SIR (See Algorithm \ref{alg:treat}). 
A simulated data set typically
has 2-7 variables selected, with the exact number varying for
different model scenarios and different simulation replicates. 
We then conduct SIR for each treatment group and obtain the first
projection direction $\hat{\beta}_{a}$, $a=1$ or
$-1$. Feature Scores are calculated for subjects in the corresponding
treatment group, either $a=1$ or $-1$, as
$u_i=\hat{\beta}_a\mathbf{x}_i$.

\begin{figure}[ptbh]
\centering
\includegraphics[width=1.0\linewidth]{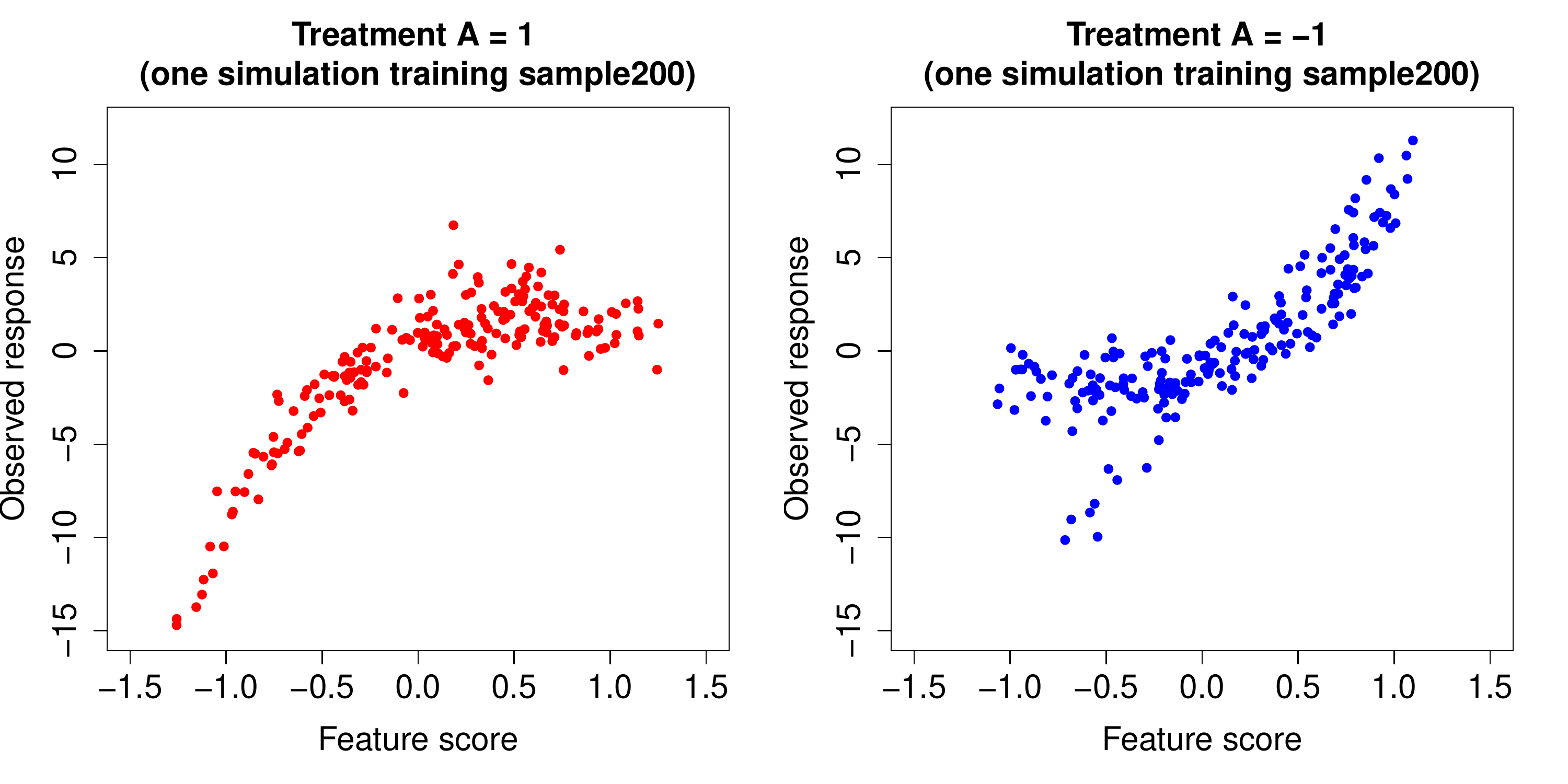}
\caption{\small{Scatter plot of $Y$ versus Feature
  Score $u_{a}=\hat{\beta}_{a}\mathbf{x}$ under each treatment (Scenario 3).} }\label{fig:obs_scatter}
\end{figure}

Figure \ref{fig:obs_scatter}
  shows two scatter plots of $y_i$ versus $u_i$, one for each
  treatment group. These plots display functional relationships
  between the response $Y$ and Feature Score and are used to predict
responses for a new observation $\mathbf{x}$. Figure \ref{fig:simplot}
shows plots of the predicted response versus Feature Scores,
where each sample $\mathbf{x}_i$, $i=1,\dots,n$, is considered as a
new observation (a test data). Each sample has two
Feature Scores, $u_{a,i}=\hat{\beta}_{a}\mathbf{x}_i$, $a=1$ or
$-1$, and two predicted treatment responses from the LOESS fits of
Figure \ref{fig:obs_scatter}. We use the R package {\tt loess()} with
its default bandwidth parameter $h=0.75$. The vertical axes of these plots use
exactly the same scale for treatment response and thus are
directly compared. It clearly demonstrates the optimal treatment option, either
$a=1$ or $-1$, for each sample. More specifically, a subject
with ID 139, as marked by a small triangle in the plots, has
the predicted response value
$\hat{Y}=-11.44723$ if it is assigned to treatment $a=1$, and 
$\hat{Y}=-2.79185$ if it is assigned to
treatment $a=-1$. This subject is then recommended to get treatment
$a=-1$ due to the larger predicted response value. 

\begin{figure}[ptbh]
\centering
\includegraphics[width=1.0\linewidth]{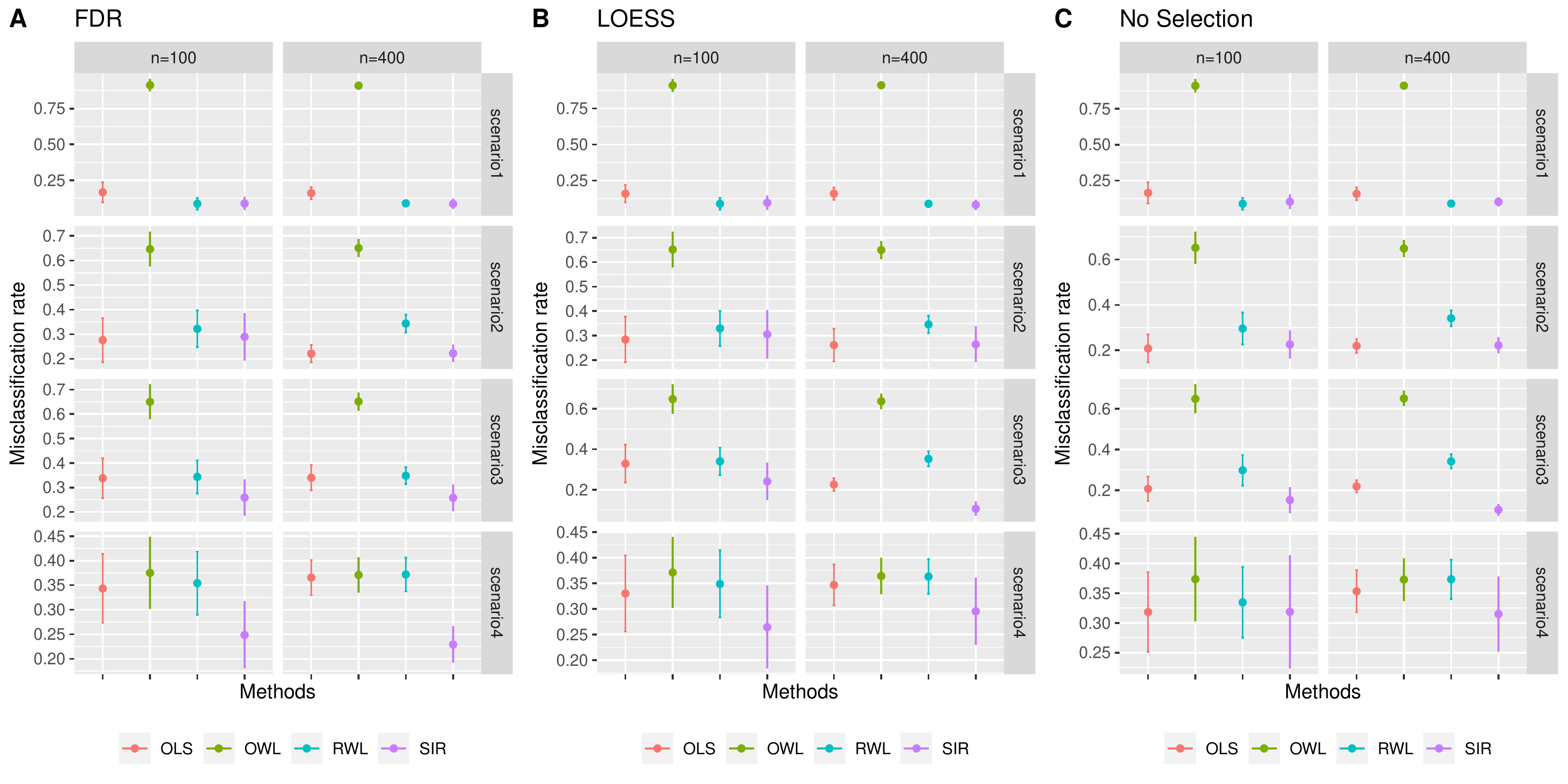}
\caption{\small{Comparison of different treatment recommendation
    methods in terms of 
  the mean (center) and the standard deviation (error bars) of misclassification rates from 1000
  simulations: A. FDR is used to screen all $p=100$ variables in the first step;
  B. LOESS is used to screen all $p=100$ variables in the first step;
  C. No screening but $p=8$.}}\label{fig:comparisons}
\end{figure}

For the simulation studies, we know the true optimal treatment recommendation,
which is the treatment option with the larger value of
$\mu(\mathbf{x})+t_0(\mathbf{x})a$, for
a given subject with covariate values $\mathbf{x}$. We are able to
evaluate our method and compare it with other existing methods by calcuating
a misclassification error. More specifically, if the treatment
recommendation through an approach gives the same treatment option as
the truth, there is no misclassification error. Otherwise, the
misclassification error is 1. Four treatment recommendation methods
are applied: Outcome Weighted Learning (OWL) \cite{Zhao12}, Residual
Weighted Learning (RWL) \cite{Zhou17}, linear regression to predict
$Y$ and then to recommend the treatment with a larger predicted value
(OLS), and our method denoted as SIR.
We use an existing R package to perform OWL and RWL, 
\newline
https://cran.r-project.org/web/packages/DynTxRegime/index.html. We
make a treatment recommendation for each  
sample while considering all other samples as the training data.
Figure \ref{fig:comparisons} displays the misclassifications
rates. The rate is the percentage of the number of misclassified
treatments 
 over the total number of patients (sample size $n$). We repeat the
 whole simulation procedure 1000 times and plot
the mean value and the standard deviation, with two error bars around the mean,
 in Figure \ref{fig:comparisons}.

In general, our approach (SIR) shows better performance with lower
misclassification rates. In particular, our 
approach performs substantially better than RWL and OWL in Scenario
2, 3, 4. The results of SIR and RWL are comparable in Scenario
1. In addition, our
approach shows lower misclassification rates than OLS in
Scenario 1, 3, 4. The results of SIR and OLS are comparable in
Scenario 2. The favorable performance of SIR is due to the general
assumption of the treatment response model, i.e., Assumption (A.2),
which gives a large approximation space for the true
conditional mean function $Q_0$. In other words, the model space of
SIR is often bigger than other existing methods with mostly linear
models. We improve the treatment recommendation by obtaining a good
estimator of $Q_0$.

\section{A case study}
We have applied our proposed method to the study of
bortezomib in treatment of multiple myeloma
\cite{Mulligan07}. Bortezomib is the
first therapeutic proteasome inhibitor tested in humans. It is
approved in the U.S. for treating relapsed multiple myeloma. As
bortezomib is a therapeutic  
choice in addition to the standard chemotherapy, there is a need to
decide which treatment should be recommended for a given patient.
Our goal is to provide a treatment
recommendation, either dexamethasone (dex) or bortezomib, based on
data information.

To achieve this, we use a genomic-clinical data set from the Gene
Expression Omnibus 
(GEO) database (GSE9782). Data from two platforms of Affymetric
microarrays (GPL96 and GPL97) are merged to obtain a large sample size,
with a total of 477 patients, 338 of them receiving
bortezomib and 139 receiving dex. The merged data contain a
smaller number of gene probesets (or simply genes) than each of the
individual platform data. On the other hand, we have verified that
significant genes from each data set are
included in the merged data. The variables considered
in our analysis include:
\begin{itemize}
  \item a set of clinical prognostic factors, i.e., gender, race, age;
  \item a treatment index, either bortezomib or dex, denoted as $A$;
  \item gene expression measurements of 168 genes in
    the merged data, denoted as $X_j$, $j=1,...,168$;
    \item clinical response denoted as $Y$ with five levels coded as 1-5
      corresponding to progressive
      disease (PD), no change (NC), minimal response (MR), partial
      response (PR), complete response (CR), respectively.
\end{itemize}
We first evaluate whether this data set provides
significant information for the prediction of treatment
response $Y$. The three clinical factors, i.e., 
gender, race, and age, have no significant effect on $Y$
($R^2=0.004179$) hence are not considered in the following
anlaysis. The Cauchy combination test \cite{Liu19} gives a p-value
$0.0004$,  
suggesting that the genomic data set contributes to the
treatment response and provides useful information for treatment
recommendation.

Given the sample size and the number of genes are comparable, 
we deem initial variable selection is necessary before
running the SIR procedure. We select a subset of 8 most significant
genes at a false discovery rate cutoff ($0.002$). They are the genes
of ribosomal proteins and translation initiation
factors. Interestingly, these genes match with the literature that
patients with perturbation of certain ribosomal proteins and
translation initiation factors showed responses to the bortezomib treatment
\cite{Mulligan07, Sulima17, Hofman17}. 
We then apply our SIR method of treatment
recommendation using this set of 8 genes and compare the
performance with OLS and RWL. For the SIR method, Feature Score is
calculated as a one-dimensional projection of the gene predictors
for each treatment group.

More specifically, we randomly
split the data into five equal-sized parts. Four parts (training data) are
used to fit a model, either OLS, SIR, or RWL, and the remaining one part
(test data) is used to evaluate the corresponding treatment
recommendation methods. Different from the simulation examples, we do
not know the true optimal treatment recommendation for this case
study hence cannot calculate the misclassification errors. Instead, we
calculate an unbiased estimator of the Value function as in 
\cite{Qian11}.  We repeat the process 1000 times and
report the mean and 
standard deviation of the estimated Value functions in Table
\ref{tab:real}. The observed treatment index $A$ in the data also
corresponds to a treatment recommendation rule. Its estimated Value
function serves as a baseline for the performance comparison.

Table \ref{tab:real} shows that SIR improves the baseline Value function
from 2.54 to 2.82 and is slightly better than the OLS and RWL
methods, although the difference from OLS and RWL is minimal.
Plots of $Y$ versus Feature Score (plots not shown here) actually
display a certain degree of linear trend and the 
predicted $Y$ curves from OLS and SIR are not very different from each
other. This explains the similar result of different methods in Table
\ref{tab:real}. On 
the other hand, RWL is computational expensive, costing about 
300 more times than SIR and OLS.

\begin{figure}[ptbh]
\centering
\includegraphics[width=1.0\linewidth]{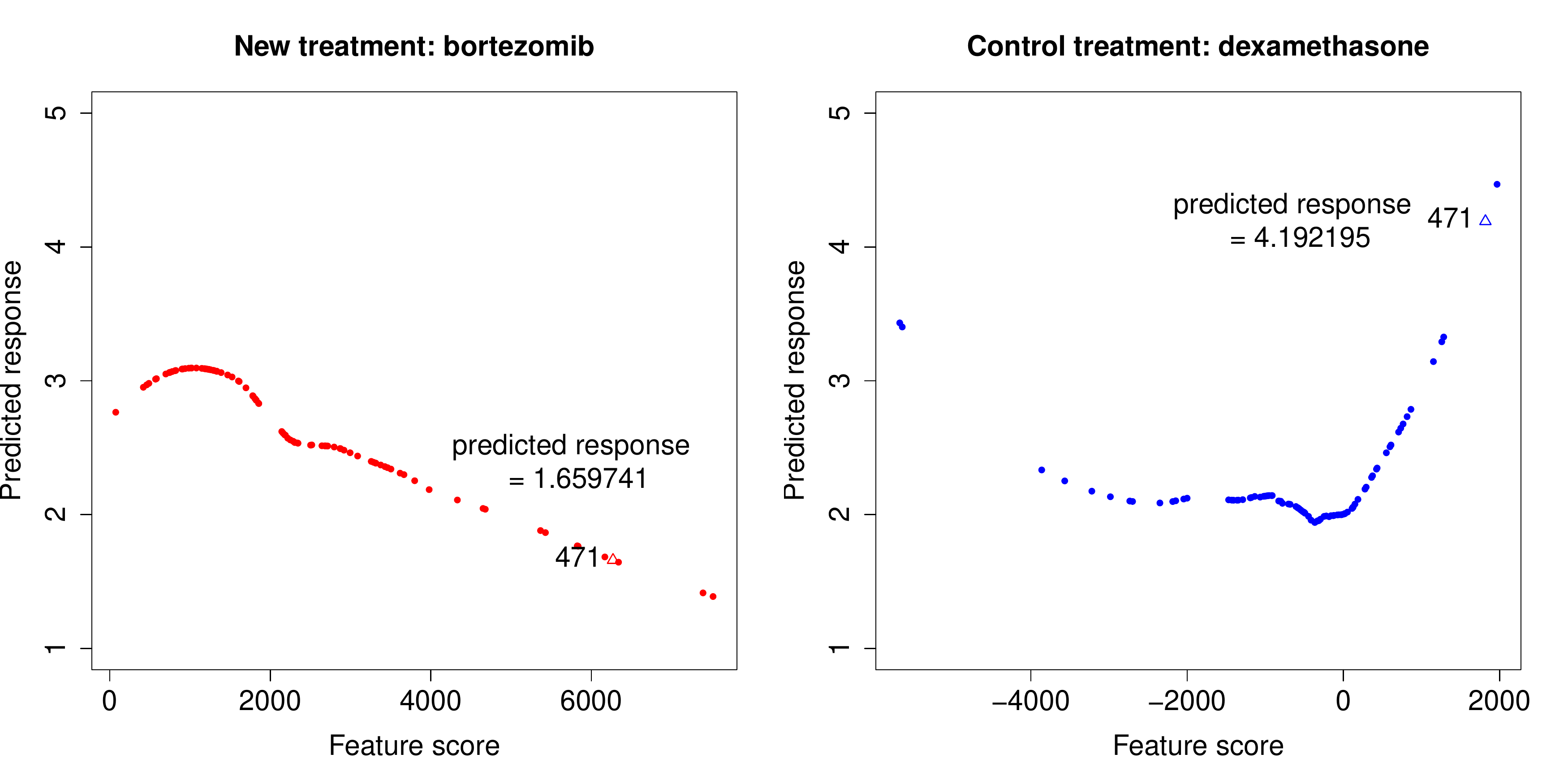}
\caption{\small{Scatter plot of treatment response versus Feature Score for
  each treatment group in a test data set of the real data example. A
  specific data point, patient ID 471, is marked for visualization of
  the optimal treatment.}}\label{fig:realplot}
\end{figure}

Figure
\ref{fig:realplot} is a plot of the predicted treatment response versus
Feature Score in a random test data set. Note that Feature Score is
different for the two treatment groups but we can still directly
compare the predicted treatment response on the vertical axes.
A specific data point, patient ID 471, is marked for visualization of
  the treatment recommendation. This patient has a lower predicted treatment
  response value under bortezomib than under dex. Therefore, the
  optimal treatment recommendation is the standard chemotherapy dex
  for this patient. This recommendation is based on the gene
  expression data through the Feature Score
  generated by SIR. To conclude, our data-guided method is able to
  provide the 
multiple myeloma patients with a treatment recommendation between
bortezomib and dexamethasone with better performance than not
used. The data-guided method attempts to connect information from the
gene expression with 
treatment responses and may reveal relationships
between genes and the corresponding 
phenotype.

\section{Discussion}
A major advantage of the proposed method lies in its low-dimensional
representation of data, i.e., the Feature Score definition, and the
automatic detection of these features through the SIR
approach. In comparison to the lasso-type approaches such as seen in
\cite{Qian11}, SIR works better than variable selection methods when the
effects from individual predictors are minimal. The features from the
SIR approach resemble the feature definition of the popular
neural network models, with a wide potential of
applications. Additionally, the SIR procedure is much  
simpler than learning a neural network model.

SIR  
is a novel method for reducing the dimension of $\mathbf X$ without going
through any model-fitting 
process in the first place. It is developed under a very general model
assumption that the treatment response $Y$ depends on the covariates
$\mathbf X$ through a low-dimensional projection space. This general
assumption corresponds to a large approximation space for the true
conditional mean function $Q_0$, hence resulting in a consistent
estimation of the optimal recommendation rule. In other words,
Assumption (A.2) is 
the most critical assumption for the theoretical guarantee, whereas
the other assumptions are standard.  

The proposed method does not, however, consider dynamic treatment regimes that
involve treatment recommendations being made at multiple times. Since
there are far 
more datasets with only one-time treatment information as compared to
multiple-time treatment information,
the proposed method would have broader applications than the methods
of dynamic treatment regimes. Besides treatment recommendation, the
proposed method can also be applied to other precision medicine research,
such as risk prediction, treatment effect estimation, and even causal
inference. Those will be the topics of our future work.

\section{Supplementary Materials}
The reader is referred to the online Supplementary Materials for
the information of data and code and technical proofs.

\bibliographystyle{plainnat}

\vspace{1cm}

\begin{table}[ptbh]
\caption{\small{Comparison of the empirical value function in a random testing dataset using different methods, OLS, SIR, and RWL. Mean (std) values of the empirical value function through 1000 resampling are reported. } } 
\label{tab:real}
\vskip5mm\centering\tabcolsep4mm
\begin{tabular}{cccc}\hline
Observed & OLS & SIR & RWL\\\hline
$2.542 (0.127)$  & $2.818 (0.157)$ & $2.825 (0.158)$ & $2.804 (0.166)$ \\\hline
\end{tabular}
\end{table}

\vspace{1cm}

\end{document}